\documentstyle[12pt,epsfig]{article}

\def\be{\begin{equation}}
\def\ee{\end{equation}}

\begin{document}

\title{Non-Equilibrium Quantum Field Theory
and \\ Entangled Commutation Relations
\thanks{To be published in  the Special Issue of Proc. of the 
Steklov Mathematical
Institute dedicated to the 90th birth day of N.N.Bogoliubov.}}
\author{
L. Accardi
\thanks{Graduate School of Polymathematics,
Nagoya University, accardi@math.nagoya-u.ac.jp}
, I.Ya. Aref'eva
\thanks{
Steklov Mathematical
Institute, Gubkin St.8, GSP-1, 117966, Moscow, Russia, arefeva@mi.ras.ru}
, I.V.Volovich
\thanks{
Steklov Mathematical
Institute, Gubkin St.8, GSP-1, 117966, Moscow, Russia, volovich@mi.ras.ru}
\\
{\it Centro Vito Volterra}
\\
{\it Universit\`a di Roma Torvergata   }}

\maketitle

\begin {abstract}

Non-equilibrium quantum field theory studies time 
dependence of processes
which are not available for the S-matrix description.
 One of  the new methods of investigation
in non-equilibrium quantum theory  is the stochastic limit method. 
This method is an extension
of the works by Bogoliubov, van Hove and Prigogine and it permits to
study not only the system but also the reservoir degrees
of freedom.
We consider the stochastic limit of translation invariant
Hamiltonians in quantum field theory and show that 
the master field satisfies a new type of commutation relations,
the so called
 entangled (or interacting) commutation relations. These relations  
extend   the  interacting Fock relations established
earlier  in non-relativistic
QED and the 
free (or Boltzmann) commutation relations which have been found 
in the large N limit of QCD .
As an application of the stochastic limit method we consider
the photon splitting cascades in magnetic field and show that photons
in cascades form entangled states ("triphons") and they
obey not Bose but a new type of statistics corresponding
to the entangled commutation relations.

\end {abstract}

\bigskip

\newpage

\setcounter{equation}{0}

\section{Introduction}

\bigskip

The basic object to study in quantum field theory is the $S$-matrix
introduced by Heisenberg. Bogoliubov and Shirkov have developed
the $S$-matrix formalism which includes all the quantities
considered in quantum field theory \cite{BS}.
The physical idea behind the $S$-matrix approach is that in the scattering
processes there exists a characteristic time scale
such that in a time regime larger 
then this time scale one can neglect 
interaction and particles evolve according to the free dynamics.

The situation in statistical physics is different because
here one has not just one but several relevant time scales
and as a result we don't have here a universal method 
comparable with the $S$-matrix approach  in quantum field theory.
One can say that the role of $S$-matix approach in non-equilibrium
statistical physics is played by various master and kinetic
equations.
It was the fundamental Bogoliubov idea about the existence of two
time scales which lead to the modern progress in the 
microscopic derivation of kinetic equations \cite{Bog1}.

Methods of quantum field theory, in particularly
Green functions, are widely used in equilibrium and 
 non-equilibrium   statistical physics \cite{BBT,AGD,Zub}.
It is well known that the so-called double-time thermodynamic
Green functions which were used by Bogoliubov and Tyablikov
\cite{BT} have had great success especially when applied to magnetic
problems. 
From the other side there are not so many
investigations in quantum field theory where results and 
methods of {\it non-equilibrium} statistical physics are exploited.  
N.N. Bogoliubov has made great
contributions  to both of these sciences.
As far as we know his works in non-equilibrium statistical 
physics and in
quantum field theory were performed completely separately. 
Probably the reason
was that important problems in quantum field theory in that 
time  were related with the scattering processes
described by the $S$-matrix and they were really very different from
typical  problems in non-equilibrium statistical 
physics such as the derivation of kinetic equations.

We would like to point out that there are important problems 
in quantum field
theory where the standard $S$-matrix description is not 
very convenient or 
even not applicable. These include not only investigation of 
bound states and spectral
problems (see \cite {BS}) but also processes with unstable  
particles \cite{Sch,GW}
(in fact almost all particles are unstable), atom-photon interactions
\cite{CDG}, elementary particles in "semidressed states"
with non-equilibrium proper fields \cite{Fei},  
electroweak baryogenesis and
phase transitions in the early Universe and in high-energy collisions
\cite{RS}, quantum optics \cite{WM} etc. In the consideration
of such processes we are interested in the time regime smaller
 than the "infinite" time when the $S$-matrix description
 becomes applicable.  
One can say that the consideration of such processes
belongs to {\it non-equilibrium quantum field theory}. One believes
 that the method of $S$-matrix 
 in quantum field theory is analogous
to the Gibbs distribution in equilibrium 
 statistical physics and that there exists
a general method (the stochastic limit, see below) in non-equilibrium
quantum field theory which provides a description of quantum phenomena
depending on time.  One of the first works
on  the systematic application of methods of non-equilibrium statistical
physics in quantum field theory is the work of Prigogine \cite{Prig1}
in which kinetic equations for the Lee model have been derived.   

A general method in non-equilibrium quantum field theory is
the method of stochastic limit \cite{AFL,ALV1}. The idea of this method is
the systematic application of the $\lambda^2t$-limit and quantum
stochastic
differential equations. One 
considers  the evolution operator $U(t)$
of quantum system for 
small coupling constant $\lambda$ and large time $t$.
Intuitively, the weak coupling -- long time limit
means that we are looking at times in which the particle has already
weakly interacted many times with the field (long time cumulative
effects). The net average effect of these 
interactions amounts to a loss of memory (Markovian approximation). 
Therefore in this limit we can expect to be 
able to approximate the microscopic time evolution, which contains 
complicated memory effects, with a simpler markovian evolution.

When we say that the $S$-matrix method is not sufficient 
in non-equilibrium quantum field theory we mean that 
the standard dynamical definition of the $S$-matrix in real time is given
for example in terms of wave operators (including dressing
 \cite{vHo1,Fad,Are1})
or LSZ-formalism. This definition can not be applied
immediately to the processes with unstable particles.
The flexible Bogoliubov--
 Shirkov approach \cite{BS} 
 to $S$-matrix in principle can be applied to the description
 of unstable particles.
  There exists a phenomenological approach
to $S$-matrix which is not based on a Hamiltonian formalism.
In this approach unstable particles are described by
the Breit-Wigner complex poles of the scattering amplitudes \cite{Che}.
The dynamical justification of this  phenomenological approach
is given in the Weisskopf-Wigner resolvent method, for a discussion
of the resolvent method see for example \cite{GW,PH}. The resolvent
method is usually used for the investigation of the degrees of freedom
of the system interacting with reservoir. The stochastic limit method
 is conveniently used for the consideration of degrees
 of freedom not only of the system but also of the reservoir.
   
 The first rigorous result about the interaction of a system 
with a reservoir where the role of a scaling limit involving $\lambda^2t$ 
begun to emerge, is due to Bogoliubov \cite{Bog2} (see the next section).  
Friedrichs, in the context of the now well known Friedrichs model 
\cite{Fri1}, was lead to consider the scaling limit
\be
\lambda\to 0,\qquad  t\rightarrow \infty ,\qquad  \lambda^2t
= \mbox {constant} 
\label{i1}
\ee 
by second order perturbation theory.

In the mid 1950's van Hove \cite{vHo1}  used 
this scaling as a device to 
extract the effects of a small perturbation of the global Hamiltonian of a 
composite system, on the reduced evolution of a subsystem and to
derive a Pauli--type master equation describing the irreversible time 
evolution of the observables of the sub--system. 
Therefore, in the quantum theory of open systems, the 
limit (\ref{i1}) is known as the van Hove  or the $\lambda^2t$ limit.

Using the perturbative development of the dynamics in powers of 
$\lambda$, van Hove argued that the terms of order $2n$ should behave as 
$\lambda^{2n}t^n$ (in contrast with the 
rough estimate $\lambda^{2n}t^{2n}$).
Furthermore he made plausible that almost all the terms of order
$2n$ should behave as $\lambda^{2n}t^{n-\varepsilon}$, for some
$\varepsilon>0$ and therefore vanish in the limit (\ref{i1}), while
the remaining terms should sum so to give a transport equation.

In the stochastic limit we study the behaviour of the system
in the large time and small coupling constant regime.
If $G(x_1,...x_n)$ is the Green function then we want
to investigate the asymptotic behaviour of
the expression $G({\bf x}_1,x_1^0/\lambda^2,...
,{\bf x}_n,x_n^0/\lambda^2)$ when $\lambda\to0$.
This can be performed by using  the anisotropic renormalization
group method \cite{BS,AAV1,Mat}.

Notice that the $\lambda^{2n}t^n$--behaviour of the terms of order $2n$ 
of the iterated series is exactly of the same order of magnitude of the 
behaviour of the moments of order $2n$ of a Brownian motion with time 
parameter rescaled as $\lambda^2 t$. Therefore {\it a posteriori} we can 
interpret van Hove's perturbative result as a fist indication that the 
limit (\ref{i1}) should evidentiate a kind of  quantum Brownian motion
or quantum white noise underlying the dynamics of the quantum system.
To describe and to  solve the dynamical equations
after the stochastic limit one has to derive the stochastic limit
for the collective operators, the so called master field. For simple models
the master field is the quantum white noise whose
creation and annihilation operators satisfy the relations
\be
\label{i2}
[b(t),b^+(t')]=\delta(t-t')
\ee

In the stochastic limit  one gets for the evolution operator
the equation
$$
\frac{dU(t)}{dt}=(F^+(t)b(t)+ b^+(t)F(t))U(t)
$$
Here the white noise operators $b(t), b^+(t)$ are
singular operator functions of $t$ and $F(t)$ is a regular 
operator function of $t$.  This singular equation
represents the Hamiltonian form of quantum stochastic
differential equations.  This equation can be explicitly
solved for many models. The crucial role for the possibility
of the explicit solution plays the commutation relation (\ref{i1}).

For more complex models the master fields are more complex
and one gets the entangled or interacting commutation relations.
The program of investigation of models of quantum field theory
in the stochastic limit consists from two parts. First we have
to find the commutation relations for the master field and
then study the singular differential equation for the evolution operator.
In this paper we shall discuss only the first part of the program.
For the consideration of the evolution operator see for example 
\cite{ALV1,AKV1,AKV2}.
 
In recent years various modifications and
deformations of the algebra of canonical commutation relations
have been discussed, see for example \cite{AV1,Gri1, AV2} and
references therein.
In particular, in the large $N$ limit
for $SU(N)$ invariant gauge theory (as well as for NxN matrix models), 
the following relations
\be
b(k)b^+(k')=\delta(k-k')\label{i3}
\ee
appear naturally \cite{AV2}. Here $k,k'$ are momentum variables. 
The algebra generated by the operators $b(k),b^+(k')$
satisfying (\ref{i3}) is called the  free (or Boltzmann) algebra.

In this paper we prove that the stochastic limit of interacting fields,
under the only constraint of momentum conservation, leads to
a new algebra of commutation relations.
 We get the relations of the form
 \be
B(p|k_1,...,k_m)B^+(p'|k'_1,...,k'_m)=n(p)\delta(E(p,k_1,...,k_m))
\delta(p-p')\delta(k_1-k_1')...\delta(k_n-k_n')
\label{i4}
\ee
where $B^\pm (p|k_1,...,k_m)$ is the master field, 
obtained as the stochastic
limit of a translation invariant interaction Hamiltonian, $n(p)$ is the
operator density of particles,
and $E(p,k_1,...,k_m)$ is the energy associated with the interaction vertex.
These relations are called the {\it entangled  (or interacting)
commutation relations}.
Notice that
the  equality (\ref{i4}) extends the Boltzmann algebra (\ref{i3}) 
because the
right hand side is an operator (in the particle space) rather than a
scalar: in this sense one speaks of {\it Hilbert module} \cite{AL1}
 (rather than
Hilbert space) commutation relations. 
$B (p|k_1,...,k_m)$ and the density $n(p)$
satisfy
\be
[n(p'),B(p|k_1,...,k_m)]=(\delta(p'-p)-\delta(p'-p+\sum k_i))
B(p|k_1,...,k_m)
\label{i5}
\ee

As one of physical applications of the above ideas
we will argue that photon splitting cascades in
the magnetic field  create entangled
states and that photons in cascades obey not Bose but
a new type of statistics --
infinite or quantum Boltzmann statistics.
Therefore,
this statistics has a physical meaning since it describes photons in
cascades and more generally the dominant diagrams in the long time/weak
coupling limit in quantum field theory.
The states in cascades are formed from
triples of entangled photons and may be called triphons. They belong to 
an interacting Fock space \cite{AL1,ALV2}. Interacting Cuntz
algebra has been considered in \cite{HS}.

The standard definition of the stochastic limit is given
as the limit of the Wightman correlation function. For
some models the limit of these correlation functions is equal to zero.
We show that the stochastic limit for the Green functions
is non-trivial even in these cases.

The paper is organized as follows.
In Sect. 2 we remind the Bogoliubov theorem
and compare it with the stochastic limit.
In Sect. 3 we discuss  the stochastic limit.
 Sections 4 and 5 are devoted to the derivation of the
  entangled commutation relations.
In Sect. 6 we discuss the stochastic limit for the Green functions
and in Sect. 7 the universality classes of the stochastic
limit and decay processes are considered. Finally Sect. 8
 is devoted to applications of the stochastic limit
to photon splitting cascades.

\section{Bogoliubov's Theorem}

 The first rigorous result, about the interaction of a system 
with a reservoir, where the role of a scaling 
limit involving $\lambda^2t$ 
begun to emerge, is due to Bogoliubov \cite{Bog2} 
 who considered a classical  
system with Hamiltonian
\be
 H=H_S+H_R+H_{int}\label{t1}
 \ee
where
\be
 H_S= {1\over 2}(p^2+\omega^2q^2), \qquad
H_R= {1\over 2}\sum_{n=1}^N(p_n^2+\omega_n^2q_n^2), \qquad\ \
H_{int}=\lambda q\sum_{n=1}^N(\alpha_nq_n)
\label{t2}
\ee
He assumed that 
$
\sum_{\sigma <\omega_n \alpha_n^2/ \omega_n^2}\to
\int_{\sigma}^{\infty}J(\nu)d\nu ~<~\infty
$
as $N \to \infty$ with $J(\nu)$ being a non-negative continuous function.
Supposing that, at $t=0$ the system is in the state
$q=q_0, p=p_0$ and that the state of the reservoir is a random
variable with the distribution
$\rho_R=\exp \{-H_R/T\}$
he proved that:
\begin{itemize}
\item
  As $N\rightarrow \infty$ the limit distribution
$\rho_S=\rho_S(t,p,q)$ of the random variables $p=p_t,q=q_t$, of the
system, exists at any time $t>0$.
\item
  If $H_S$ is given by (2), define the function
$\rho_S^0=\rho_0(t,H_S)=\rho_0(t,p,q)$, by:
$$\rho_S^0(t,H_S)= {\omega\over 4\pi^2T(1-e^{-2a\lambda^2t})}
\int_0^{2\pi}d\phi\exp\{- {H_S+E_0
e^{-2a\lambda^2t} -2\sqrt{H_SE_0}e^{-2a\lambda^2t}\cos\phi\over
T(1-e^{-2a\lambda^2t})}\}$$
where $ a= {\pi\over 4}J(\omega)$ and $E_0$ is
$ (p_0^2+ q_0^2) /2$
where $(p_0,q_0)$ is the initial state. 
Then for small $\lambda$ the
function $\rho_S$ can be approximated by the function $\rho_S^0$, in the
sense that, for any positive $\alpha, \beta$ as $\lambda\rightarrow 0$,
one has, uniformly in the interval
$ {\alpha\over \lambda^2}<t< {\beta\over \lambda^2}$:
$$
 {1\over \Delta t_{\lambda}}\int_t^{t+\Delta t_{\lambda}}
(\rho_S-\rho_S^0)dt\rightarrow 0 $$
for any subsequence $\{\Delta t_{\lambda}\}$  such that
$\lambda^2\Delta t_{\lambda}\rightarrow 0, ~~
\Delta t_{\lambda}\rightarrow \infty$
\end{itemize}

He gave a rigorous proof of this result using a Volterra 
integro-differential equation.

Bogoliubov's condition $ \lambda^2\Delta t_{\lambda}\rightarrow 0$ 
is 
different from the one used in the stochastic limit 
($ \lambda^2 t \rightarrow $ constant), notice however 
that $\rho_S^0(t)$ depends only on $\lambda^2 t$ and also
that for $t\to\infty$ one gets the Gibbs distribution,
$$
\rho_S^0(t)\to\frac{\omega}{2\pi T}e^{-H_S/T}
$$

\section{The Stochastic Limit}

The stochastic limit is now widely used in the
consideration of the long time/weak coupling behaviour of 
quantum dynamical
systems with dissipation, see for example \cite{ALV1}.

Let be given a quantum system described by the Hamiltonian
$$
H=H_0+\lambda V
$$
where $\lambda$ is the coupling constant.
The starting point of the stochastic limit is the equation for the
evolution operator in interaction picture
$${dU^{(\lambda)}(t)\over dt}\,=-i\lambda V(t)U^{(\lambda)}(t)$$
where
$$V(t)=e^{itH_0}Ve^{-itH_0}
$$
The main idea is that there exist a new quantum field (master field) and a
new evolution operator ${\cal U}(t)$ (they both live on a space different
from the original one) which approximates the old one
$$U^{(\lambda)}(t)\approx{\cal U}(\lambda^2t)$$
and the approximation is meant in the sense of appropriately chosen
matrix elements. The above approximation suggests a natural
interpretation of the van Hove rescaling \cite {vHo1}
$\lambda\to0$, $t\to\infty$
so that $\lambda^2t\sim\hbox{constant }=\tau$ (new time scale): it means
that we measure time in units of $1/\lambda^2$ where $\lambda$ measures the
strength of the self--interaction. By putting $\tau =1$ we see that the van Hove
rescaling is equivalent to the time rescaling $t\to t/\lambda^2$, and
therefore the limit $\lambda\to0$ will capture the dominating
contributions to the dynamics in the new time scale (the error can be
estimated to be of order $\lambda^2$).
It is remarkable that in this limit the dominating contributions can
be explicitly resummed giving rise to a new unitary operator.

A simple change of variables
 shows that the time rescaling $t\to t/\lambda^2$
is equivalent to the following rescaling of the Schr\"odinger equation
for the evolution operator:
\be
{dU^{(\lambda)}(t/\lambda^2)\over dt}\,=
-i{1\over \lambda} V(t/\lambda^2)U^{(\lambda)}(t/\lambda^2)
\ee
The unitary operator ${\cal U}(t)$ is
then obtained by taking the limit $\lambda\to0$:
\be
{\cal U}(t)=\lim_{\lambda\to0}
U^{(\lambda)}(t/\lambda^2)
\ee
 and the corresponding limit equation is
$${d\,{\cal U}(t)\over dt}\,=-i{\cal V}(t){\cal U}(t)$$
where
$${\cal V}(t)=\lim_{\lambda\to0}{1\over\lambda}\,V\left({t\over
\lambda^2}\right)$$

For a number important models the interaction Hamiltonian
has the form
$$
V={\cal A}+{\cal A}^+
$$
where ${\cal A}$ is a monomial in the creation and annihilation
operators.
 The master field is given by the asymptotic behaviour of the collective
operator $${\cal A}_{\lambda}(t)
= \frac{1}{\lambda}
{\cal A}(\frac{t}{\lambda ^{2}})$$ and its Hermitian conjugate.
Here $${\cal A}(t)=e^{itH_0}{\cal A}e^{-itH_0}$$.

The stochastic  limit is meant in the sense of the 
convergence of correlation functions $~~~~~$
$<{\cal A}^{\epsilon_1}
(t_{1}/\lambda)...{\cal A}^{\epsilon_n}
(t_{n}/\lambda)>$. Here $\epsilon_i=\pm$.

\section{
Entangled Commutation Relations
}

In recent years there has been an interest in various modifications and
deformations of the algebra of canonical commutation relations.
Especially the  simple relations  (\ref{i3}) were discussed. An
extension of the algebra (\ref{i3}) has been given in \cite{ALV2} as the
algebra describing the interacting Fock space \cite{AL1} obtained in the
stochastic limit for non--relativistic QED.

In this paper we prove that the stochastic limit of interacting fields,
under the only constraint of momentum conservation leads to
a generalization of the algebra (\ref{i3}). We
obtain that the new algebra has as its generators the
master field $B(p,k)$
depending  on two momenta $p$ and $k$ and the operator density of 
particles $n(k)$ which
satisfy the relations

\be
B(p,k)B^+(p',k')=n(p)\delta(E(p,k))\delta(p-p')\delta(k-k'),
\label{e.3}
\ee

\be
[n(p'), B(p,k)]=
(\delta(p'-p))-\delta(p'-p+ k))
B(p,k),
\label{e.4}
\ee

\be
[n(p),n(p')]=0
\label{e.5}
\ee
Here  $E(p,k)$ is the energy associated to the interaction vertex.

We call the relations (\ref{e.3})-(\ref{e.5}) 
{\it entangled commutation relations}  because, on one hand they
allow to calculate correlations of any order among the field and, on the
other hand they show that the master fields are  not kinematically
independent.

In the construction of these operators we use the Van Hove time rescaling
$t\to t/\lambda^2$, $\lambda\to0$ where
$\lambda$ is the coupling  constant. 
\footnote{To avoid a discussion of renormalization procedure
we assume that there is an ultra-violet cut-off}.

One can get a generalization of the algebra 
(\ref{e.3})-(\ref{e.5})
for the multiparticle master 
field $B(p|k_1,...,k_n)$.

In \cite{ALV2} a generalization of the algebra (\ref{i3}) of the
following form has been obtained:

\be
B(k)B^+(k')=\delta(k-k')\delta(\omega(k)+\hat{P}\cdot k)\ ,\ \
[\hat P,B(k)]=kB(k)\label{i2}
\ee
with  $\hat{P}$ being  the operator  of momenta of particles.
This algebra is not realized in the usual Fock space but in the
interacting Fock space. One can get the algebra (\ref {i2})
from the algebra (\ref{e.3})-(\ref{e.5}) if we set
$$
\hat P=\int pn(p)dp,~B(k)=\int B(p,k)dp
$$

These relations  show
that, contrarily to what happens before the limit, the observables
of the particle do not commute with the master field. In other words:
before the stochastic limit the particle and the field are kinematically
independent but nonlinearly related by the dynamics; after the stochastic
limit the dynamics is simplified but  particle and field are no longer
kinematically independent: this is what we call {\it entanglement}.
These new features imply that the algebra (\ref{e.3})-(\ref{e.5}) 
is not realized in the
usual Fock space but in the interacting Fock space \cite{AL1}.

Let us  consider the Hamiltonian

$$H_\lambda=H_0+\lambda V$$
where 
the  free Hamiltonian
$$
H_0=\int\varepsilon(p)c^+(p)c(p)d^3p+\int\omega(k)a^+(k)a(k)d^3k,
$$

$$\{c(p),c^+(p')\}=\delta(p-p'),~[a(k),a^+(k')]=\delta(k-k')$$
and the  interaction Hamiltonian:
\be
V=\int d^3kd^3pg(k,p)(c^+(p)c(p-k)a(k)+h.c.)
\label{intham}
\ee
Here $g(k,p)$ is a test function and $\varepsilon(p)$ and
$\omega(k)$ are one-particle dispersion laws, for example
$\varepsilon(p)=p^2/2,~\omega(k)=|k|$.

The  rescaled collective fields  in this case have the form
\be
{\cal A}_\lambda(p,k,t)={1\over\lambda}\,
e^{{itH_{0}\over\lambda^2}}c^+(p)a(k)c(p-k)
e^{-{itH_{0}\over\lambda^2}}=
{1\over\lambda}\,c^+(p)a(k)c(p-k)e^{itE(p,k)
/\lambda^2}
\label{1.1}
\ee

\be
{\cal A}^+_\lambda(p,k,t)=
{1\over\lambda}\,e^{{itH_{0}\over\lambda^2}}
c^+(p-k)a^+(k)c(p)e^{-{itH_{0}\over\lambda^2}}=
{1\over\lambda}\,
c^+(p-k)a^+(k)c(p)e^{-itE
(p,k)/\lambda^2}
\label{1.1'}
\ee
where
\be
E(p,k)=\epsilon (p)-\omega(k)-\epsilon (p-k)
\label{en}
\ee
is the corresponding energy. One has the following main theorem.

\bigskip

{\bf Theorem 1.} {\it The stochastic limit
$$
\lim_{\lambda\to0}{\cal A}_\lambda(p,k,t)=B^-(p,k,t),~~
\lim_{\lambda\to0}{\cal A}^+_\lambda(p,k,t)= B^+(p,k,t)
$$
exists in the sense of the convergence of the matrix elements
( $\epsilon_i=\pm$)
$$
\lim_{\lambda\to0}<0|c(q){\cal A}_\lambda^{\epsilon_1}(p_1,k_1,t_1)...
{\cal A}_\lambda^{\epsilon_n}(p_n,k_n,t_n)c^+(q')|0>
=
$$
\be
\label{1.1a}
(\Psi_0,c(q)B^{\epsilon_1}(p_1,k_1,t_1)...
B^{\epsilon_n}(p_n,k_n,t_n)c^+(q')\Psi_0)
\ee
as distributions  and the limiting
operators $B^-= B$ and  $B^+$ satisfy the 
 entangled commutation relations
\be
B(p,k,t) B^+(p',k',t')=2\pi\delta(t-t')\delta(p-p')\delta(k-k')
\cdot\delta(E(p,k))n(p)\label{1.3}
\ee

\be
[n(p'), B^{\mp}(p,k,t)]=(\pm)(\delta(p'-p)
-\delta(p'-p+ k)) B^{\mp}(p,k,t)\label{23}
\ee

\be
[n(p),n(p')]=0
\label{23a}
\ee
Here
$\Psi_0$ is the vacuum in the new Hilbert space, 
$B(p,k,t)\prod _i c^+(q_i)\Psi_0 =0$.
We use the same notations
for the creation and annihilation operators of $c$-
particles in the original and in the new Hilbert spaces.
$n(p)$ 
is   the operator density of the $c$-particles,
$n(p)=c^+(p)c(p)$}.

If we set
$$
B(p,k,t)=b(t)\otimes B(p,k)
$$
where
$$
b(t)b^+(t')=2\pi \delta (t-t')
$$
then we get the relations (\ref{e.3})
$$
B(p,k)B^+(p',k')=n(p)\delta(E(p,k))\delta(p-p')\delta(k-k')
$$

The theorem will be proved in the next section.
Note that to get non-zero in the RHS of (\ref{1.3})
we have to chose  suitable dispersion relations, so that there are 
non-trivial solutions of equation $E(p,k)=0$.
We will discuss this point in Sect.6.

\section{ Proof of the Theorem}

\bigskip
Let us consider the matrix elements

\be
\langle0|c(q)\prod^v_{i=1}
{\cal A}_{\lambda}^{\epsilon_i}(p_i,k_i,t_i)c^+(q')|0\rangle
\label{tg.1}
\ee
To evaluate (\ref{tg.1}) we apply Wick's theorem \cite{BS}
and consider only the corresponding connected diagrams. Each vertex
contains $3$ lines  which are characterized by two momenta $(k_i,p_i)$.
We find the momentum corresponding to the $3$rd line using the momentum
conservation. If a diagram contains $L$ loops then only $L+A+B-1$
momenta are independent ($L$ momenta for loop variables and $A+B-1$
momenta for the exterior lines).

For every vertex there is the corresponding  energy exponent. These
energies depend on the momenta of the lines that enter to the given vertex,
$$E^\pm_i=E^\pm_i(k_i,p_i)$$
via dispersion laws,
\be
E^{\pm}_i=[\epsilon (p_i)
-\epsilon (p_i\pm k_i) \pm  \omega(k_i)]
\label{34}
\ee

In the proof of the theorem we will use the following lemma.

\bigskip

{\bf Lemma 1}. {\it  One has the following relation 
in the sense of distributions}
$$
\lim_{\lambda\to 0}\frac{1}{\lambda^2}e^{itE(p,k)/\lambda^2}=
2\pi\delta(t)\delta(E(p,k))
$$
\bigskip

The proof of the lemma is standard, see \cite{Vla1}.
\bigskip

Let us consider a connected diagram corresponding to the matrix element
(\ref{tg.1}).
The proof of the theorem consists of three parts. First we
prove that only diagrams consisting of pairs of conjugated
vertices don't vanish in the limit $\lambda\to 0$. Next we prove
that such diagrams are in fact non-crossing or half-planar diagrams.
And finally we show that the non-crossing diagrams are described
by the entangled commutation relations.

Generally, the sets of momenta corresponding to different vertices are
different. However,  it may happens that
{\it the same set of momenta corresponds to two different vertices}.
More precisely, momenta which come in the first vertex come out from
the second one and viceversa.
\bigskip

{\bf Definition 1}. We say that two incident vertices of a given
connected diagram are {\it conjugated}  if the momenta coming in
the first vertex come out from the second vertex, i.e.
the vertices have the same momenta but with the opposite orientation.
\bigskip

If the $i$--vertex has a conjugated vertex then we denote the latter by
$\hat i$.
A typical example of diagrams containing at least one pair of conjugater
vertices is a diagram with a mass insertion such that this
insertion contains a line that does not cross others lines of the
diagram (see Fig.1).

\begin{figure}
\begin{center}
\epsfig{file=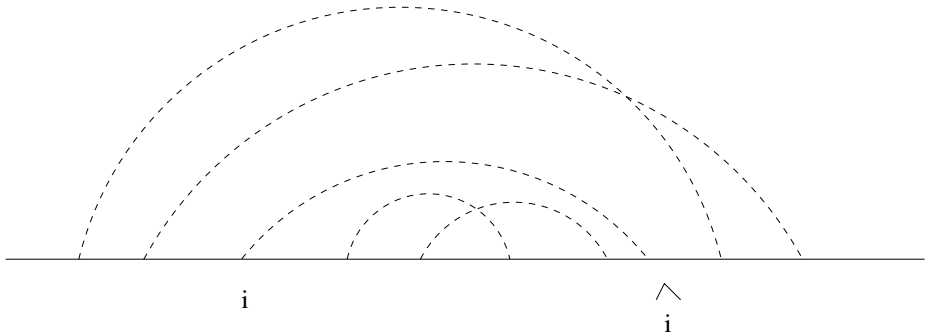,
   width=350pt,
  angle=0
 }
\end{center}
\label{Fig1}
\caption{Diagram with a pair of conjugated vertices}
\end{figure}
Let us prove the main
\bigskip

{\bf Lemma 2}.  {\it If a connected  
diagram doesn't  consist   only from pairs of conjugated
vertices then it vanishes in the limit
$\lambda\to0$ (in the sense of distributions)}.
\bigskip

{\it Proof\/}. To a given diagram, representing a matrix element
(\ref{tg.1}) being integrated over $t_1,\dots,t_v$ with test functions,
corresponds the  expression that schematically can be written
as

\be
{1\over\lambda^v}\,\int e^{i\,\sum^v_{i=1}E_it_i/\lambda^2}
\phi(t,p,q)\prod^{A+B-1}_adp_a\prod^L_{l=1}dq_l\prod^v_{i=1}
dt_i\label{tg.3}
\ee
here by $t$ we mean $t_1,\dots,t_v$, by $p$ we mean $p_1,\dots,p_A$,
$p'_1,\dots,p'_B$ and $q$ denotes the set of independent momenta associated
with the diagram under consideration.
$E_i$ are given by (\ref{34}) and
$\phi(t,p,q)$ is a  test function.

To evaluate the asymptotic behaviour of this expression 
when $\lambda\to0$ we will
 make the change of variables corresponding to the conjugated vertices.
 Notice that for the diagram doesn't vanish, the number of vertices $v$
 should be even, $v=2n$. Suppose that there are
 $n_0$ pairs of conjugated vertices which are denoted $\{i_1,\hat i_1,
 ...,i_{n_0},\hat i_{n_0}\}$.
 Let us divide the set of all vertices $\{1,...,2n\}$
 into two disjoint subsets $\{i_1, i_2,\dots,i_{n_0}, i_{n_0+1},
 ...,i_n\}$
 and $\{\hat i_1, \hat i_2,
 \dots,\hat i_{n_0}, i_{n+n_0+1},\dots i_{2n}\}$
 such that in every subset there are no conjugated vertices.
 We denote the corresponding set of time variables 
$\{t_{i_1},...t_{i_n}\}$ 
by $t^{(1)}$ and the set 
$\{t_{\hat i_1},...t_{\hat i_{n_0}},t_{i_{n+n_0+1}},...t_{i_{2n}}\}$ 
by $t^{(2)}$.

Now  we perform the following change of variables

\be
\label{36a}
(t^{(1)},t^{(2)}) \to (\tau, t^{(2)})
\ee
$$
t^{(1)}=t^{(2)}+\lambda ^2 \tau ,
$$
or more precisely 
\be
(t^{(1)},t^{(2)})=
(t_1,\dots,t_{2n})
\to
(\tau, t^{(2)})=
(\tau_1,\dots,\tau_n;\ t_{\hat i_j},
j=1\dots,n_0;\ t_{i_{n+r}},\ r=n_0+1,\dots,n)
\ee
\be
t_{i_j}=t_{\hat i_j}+\lambda^2\tau_j\ ,\quad1\leq j\leq n_0\label{tg.4}
\ee
\be
t_{i_j}=t_{i_{n+j}}+\lambda^2 \tau_j\ ,\quad n_0<j\leq n
\ee
After this the integral (\ref{tg.3}) takes the following form

$$
\int e^{i\sum^n_{j=1}\tau_jE_{i_j}}e^{i\sum^{n_0}_{j=1}(E_{i_j}+
E_{\hat i_j})t_{\hat i_j}/\lambda^2}\cdot
e^{i\sum^n_{j=n_0+1}(E_{i_j}+E_{i_{n+j}})t_{i_{n+j}}/\lambda^2}
$$

\be
\label{tg.5a}
\phi(t^{(2)}+\lambda^2\tau,t^{(2)},p,q)
\cdot\prod^n_{j=1}d\tau_j\prod^{n_0}_{j=1}dt_{\hat i_j}\prod^n_{j=
n_0+1}dt_{i_{n+j}}\prod dp\cdot\prod dq
\ee
By definition of conjugated vertices 
$E_{i_j}+E_{\hat i_j}=0$ and we left with 

$$
\int e^{i\sum^n_{j=1}\tau_jE_{i_j}}
e^{i\sum^n_{j=n_0+1}(E_{i_j}+E_{i_{n+j}})t_{i_{n+j}}/\lambda^2}
$$

\be
\label{tg.5}
\phi(t^{(2)}+\lambda^2\tau,t^{(2)},p,q)
\cdot\prod^n_{j=1}d\tau_j\prod^{n_0}_{j=1}dt_{\hat i_j}\prod^n_{j=
n_0+1}dt_{i_{n+j}}\prod dp\cdot\prod dq
\ee
Here $t^{(2)}+\lambda^2\tau$ 
schematically represents the dependence of the half of
the $t$--variables on $\lambda$. When $\lambda\to0$ we can neglect the
dependence of $\phi$ on $\lambda$ and the integration over $\tau$  gives
the product of $\delta (E_{i_j})$

\be
\int\prod^n_{j=1}\delta(E_{i_j})e^{i\sum^n_{j=n_0+1}(E_{i_j}+
E_{i_{n+j}})t_{i_{n+j}}/\lambda^2}
\phi(t,t,p,q)\prod^n_{j=n_0+1}dt_{i_{n+j}}\prod dp\prod dq\label{tg.6}
\ee

Note that the second exponent in the expression ({\ref{tg.5})
vanishes since the energies in the conjugated vertices are equal.

Suppose that in the diagram there are
non-conjugated vertices, i.e. $n\not=n_0$. When $\lambda\to0$ 
the expression (\ref{tg.6})
goes to zero since,
according to our assumption the set of momenta in vertices $i_j$ and
$i_{n+j}$, $n_0<j\leq n$ do not coincide and therefore the functions
$E_{i_j}+E_{i_{n+j}}$ as functions of momenta don't vanish and
therefore according the Riemann-Lebesgue lemma we get zero in the limit
$\lambda\to0$.

In the case when $n=n_0$ the exponent in (\ref{tg.6})
vanishes and generally we get the non--zero answer:
\be
\int\left(\prod^n_{i=1}\delta(E_{i_j})\right)\phi(\{t_{i_j}\},
\{t_{i_j}\},p,q)\prod dp\prod dq\label{tg.7}
\ee

The lemma is proved.
\bigskip

{\bf Lemma 3.} {\it If a connected diagram with two
external lines consists only from
pairs of conjugated vertices then it is half-planar,
i.e. it can be drown in the half-plane without self intersections.}

\bigskip

We will not present here the simple proof of this lemma.

\bigskip

{\bf Lemma 4.} {\it  The limiting expression (\ref{tg.7})
is equal to the RHS of the relation (\ref{1.1a}).} 

\bigskip

The proof is obtained by using the algebra (\ref{1.3}), (\ref{23a}).
\bigskip

Now the theorem follows from the above four lemma.
\bigskip

{\bf Remark 1.} We have considered only matrix
elements with two external particles and as a result 
the matrix element doesn't vanish only if $v$ is even, $v=2n$. In the 
general case of  matrix elements with an arbitrary
number of external lines the number of vertices $v$ can be odd.
 For the case of odd $v$, $v=2n-1$, we once again select $n$ vertices
and make the following change of variables

\be
(t_1,\dots,t_{2n-1})\to(\tau_1,\dots,\tau_n;t_{\hat i_j},j=1,\dots,
n_0;t_{i_{n+r}},r=1,\dots,n-1-n_0)
\ee
with the $\tau_i$ as before (see (\ref{tg.4})).
$$t_{i_j}=t_{\hat i_j}+\lambda^2\tau_j\ ,\quad1\leq j\leq n_0$$
$$t_{i_j}=t_{i_{n+j}}+\lambda^2\tau_j\quad n_0+1\leq j\leq n-1$$
$$t_{i_n}=\lambda^2\tau_n$$
The difference with the case of even $v$ is that we get an extra factor
$\lambda$ in front of the integral
$$\lambda\int e^{i\sum^n_{j=1}\tau_jE_{i_j}}e^{i\sum^{n_0}_{j=1}
(E_{i_j}+E_{\hat i_j})t_{\hat i_j}/\lambda^2}
e^{i\sum^{n-1}_{j=n_0+1}(E_{i_j}+E_{i_{n+j}})t_{i_{n+j}}/\lambda^2}$$

\be
\phi(t+\lambda\tau,t,p,q)\prod dpdq\prod^n_{j=1}dt_j
\prod^{n_0}_{j=n_0}dt_{i_{n+j}}\label{8}
\ee
Here we use the same schematical notations as in (5). When $\lambda\to0$ we
neglect the $\lambda$--dependence of $\phi$ and we get a product of
$\delta$--functions. The second exponent goes out. The third exponent
disappears in the case then $n_0=n-1$, but since we have an extra
factor $\lambda$ the expression (8) always goes to zero as $\lambda\to0$.
\bigskip

{\bf Remark 2}. The theorem admits a generalization to the case
of an arbitrary number of external particles

$$<0|\prod_ic(q_i){\cal A}_
\lambda^{\epsilon_1}(p_1,k_1,t_1)...
{\cal A}_\lambda^{\epsilon_n}(p_n,k_n,t_n)\prod_jc^+(q_j')|0>
$$

{\bf Remark 3}.  For the composite operators
$$
{\cal A}_\lambda(t,p,k_1,...,k_m)= {1\over\lambda}\,c^+(p)c(p-k_1)
a^{\epsilon_1}(k_1)...a^{\epsilon_m}(k_m)e^{itE(p,k_1,...,k_m)
/\lambda^2}
$$
in the stochastic limit one gets the entangled commutation relations
(\ref{i4}).
\bigskip
\bigskip
\bigskip
\bigskip
\bigskip

\section {The stochastic limit for Green functions}

The master field in 
the standard formulation of the stochastic 
limit is defined by the limit 
at $\lambda\to0$ of the Wightman correlation functions
\be
<0|A^{(1)}_\lambda(t_1)...A^{(n)}_\lambda(t_n)|0>
\label{two.1}
\ee
This limit defines the Hilbert space in which the master
field lives. For some models the limit of these correlation functions
is trivial (equal to zero). However this doesn't mean that
the stochastic limit of such the model is trivial because
we can consider another
 natural formulation
of the stochastic limit 
which is defined by the convergence of the chronologically ordered
correlation functions (Green functions)
\be
<0|T(A^{(1)}_\lambda(t_1)...A^{(n)}_\lambda(t_n))|0>
\label{two.2}
\ee
For the evolution operator one has to consider
\be
<0|T(A^{(1)}_\lambda(t_1)...A^{(n)}_\lambda(t_n)U(t))|0>
\label{two.3}
\ee
Here the $T$-product is defined as
$$
T(A^{(1)}_\lambda(t_1)...A^{(n)}_\lambda(t_n))=
A^{(i_1)}_\lambda(t_{i_1})...A^{(i_n)}_\lambda(t_{i_n})
$$
if $t_{i_1}\geq...\geq t_{i_n}$.

The limit of the Green functions can be nontrivial even if
the limit of the Wightman functions vanishes. In the simplest case
in the first formulation we use the relation
\be
\lim_{\lambda\to0}\frac{1}{\lambda^2}e^{itx/\lambda^2}=2\pi
\delta(t)\delta(x)
\label{two.4}
\ee
 and in the second
 \be
\lim_{\lambda\to0}\frac{1}{\lambda^2}\theta(t)e^{itx/\lambda^2}=
i\delta(t)\frac{1}{x+i0}
\label{two.5}
\ee
Here $\theta(t)=1,t\geq0;~\theta(t)=0,t<0$.
In particular if
$$
A_\lambda(t,p)=\frac{1}{\lambda}e^{it\omega(p)/\lambda^2}a(p)
$$
where $[a(p),a^+(p')]=\delta(p-p'),~p,p'\in R^3$ then
the stochastic limit of the Wightman functions is
$$
<0|A_\lambda(t,p)A^+_\lambda(t',p')|0>=\frac{1}{\lambda^2}
e^{i(t-t')\omega(p)/\lambda^2}\delta(p-p')\to2\pi\delta(t-t')
\delta(\omega(p))\delta(p-p')
$$
and the stochastic limit of the Green functions is
$$
<0|T(A_\lambda(t,p)A^+_\lambda(t',p'))|0>=\frac{1}{\lambda^2}
\theta(t-t')e^{i(t-t')\omega(p)/\lambda^2}\delta(p-p')\to i\delta(t-t')
\frac{1}{\omega(p)+i0}\delta(p-p')
$$
It is important to notice that in the last formula one gets 
a nontrivial limit even if $\delta(\omega(p))=0$
when in the former formulation the limit is trivial.

\section {Stochastic Limit and Decay}

In the formulation  of  Theorem 1 we assumed the special form of the 
interaction Hamiltonian
 (\ref{intham}). In local quantum field theory
 the typical Hamiltonian is more complicated than (\ref{intham}).
 In particular, for the Yukawa interaction of fields $\psi$
 and $\phi$ the Hamiltonian has the form
$$H_\lambda=H_0+\lambda V$$
where $H_0$ is the sum of the free Hamiltonians for the fermionic field
$\psi$ and for the scalar field $\phi$ 
with relativistic dispersion laws,
\be
\label{d1}
\omega_a(k)=\sqrt{m_a^2+k^2},~~~~~\omega_b(k)=\sqrt{m_b^2+k^2}
\ee
and

$$V=\int d^3xg\overline\psi\psi \phi=$$
\be
\int d^3kd^3pg(k,p)(c^+(p)c(p-k)a(k)+c^+(p)c^+(k-p)a(k)+
c^+(p)a^+(k)c^+(-p-k)+h.c.)\label{q2}
\ee
Here $g(k,p)$ is a test function. 
This expression is not a well defined operator in the Fock space but it
defines a bilinear form. We have the following collective operators
\be
{1\over\lambda}\,c^+(p)c(p-k)a(k)\exp{it\over\lambda^2}\,(\varepsilon(p)
-\varepsilon(p-k)-\omega(k))
\label{q3}
\ee
\be
{1\over\lambda}\,c^+(p)c^+(k)a(p+k)\exp{it\over\lambda^2}\,
(\varepsilon(p)+\varepsilon(k)-\omega(p+k))
\label{q4}
\ee
\be
{1\over\lambda}\,c^+(p)c^+(k)a^+(-p-k)\exp{it\over\lambda^2}\,
(\varepsilon(p)+\varepsilon(k)+\omega(p+k))
\label{q5}
\ee

In the limit $\lambda\to0$ the operator (\ref{q5}) 
vanishes 
because in the correlation functions one gets

$$\delta(\omega_c(p)+\omega_a(k)+\omega_c(p+k))=0$$
due to the positivity of energy.

The limit of the operator (\ref{q3}) is also zero since
for relativistic dispersion laws (\ref{d1})
$$\delta(\omega_c(p)+\omega_a(k)-\omega_c(p+k))=0$$
Only operator (\ref{q4}) has a chance to be non-zero.
In this case we have to find non-trivial solutions of
\be
\omega_c(p)+\omega_c(k-p)=\omega_a(k)
\label{dc}\ee
There are solutions if $$m^2_a>2m^2_c,$$ 
i.e. if one has the decay.
 Therefore we obtain that the relativistic
interaction Hamiltonian (\ref{q2}) is within the same stochastic {\it
universality class\/} as the  interaction Hamiltonian (\ref{intham})
\be
V=\int d^3kd^3pg(k,p)(c^+(p)c^+(k-p)a(k)+h.c.)
\ee

In quantum electrodynamics (QED) the interaction has the form
$$V=\int d^3x\overline\psi\gamma^\mu\psi A_\mu$$
If one   neglects  the spinor and polarization indices 
then the free Hamiltonian has the form (\ref{d1})
and the  interaction Hamiltonian has the form (\ref{q2}).
We see that the
stochastic limit of QED Hamiltonian is reduced to the
form of the above discussed Hamiltonian and as a result we obtain the
trivial limit, since
there is no decay in the standard QED. However, if we consider
QED in external field we can get a non-zero result due to the change of
dispersion law.
One of such examples we will consider in the next section.

\section{Photon splitting cascades  and a new statistics}
In the previous section we have seen that the stochastic limit of Wightman
functions
in relativistic QED is trivial  and that 
it can be non-trivial in the presence of external fields.
It is known that there is the splitting of a
photon into two in an external magnetic field. This splitting is 
one of the most interesting manifestations of the non--linearity of
Maxwell's equations with radiative corrections in
an external magnetic field. In a constant uniform
field, this process occurs with conservation of energy and momentum. The
process was considered by Adler {\it et al.} in the early '70s by using
the Heisenberg--Euler effective Lagrangian \cite{1,2,3}.
Photon splitting was considered as a possible mechanism for the production
of linearly polarized gamma--rays in a pulsar field. Recently the splitting
of photons has found astrophysical applications in 
the study of annihilation
line suppression in gamma ray pulsars and spectral formation of gamma ray
bursts from neutron stars \cite{7,8}.  Photon splitting cascades have also
been used  in models of soft gamma-ray repeaters, where they soften
the photon spectrum \cite{15,14}. The process of photon splitting is
potentially important in applications to a possible explanation of the 
origin
of high energy cosmic rays from Active Galactic Nuclei \cite{24}. A
recalculation of the amplitude for photon splitting in a strong magnetic
field has been performed recently in \cite{4,5,6}.

We will start from a discussion of  the theory of photon splitting
cascades and show the emergence of infinite statistics in this theory
and then discuss its connection with the
stochastic limit of quantum field theory.
In the decay of a photon with momentum $k$ into photons with
momentum $k_1$ and $k_2$, we have conservation of momentum
and energy $k=k_1+k_2\ ,$
$\omega(k)=\omega_1(k_1)+\omega_2(k_2)$.
For photons in vacuum, in the absence of external fields,
$\omega=\omega_1=\omega_2=k$ and although these two equations have a 
solution
the decay is forbidden by the invariance under charge
conjugation (Furry's theorem).

In a constant uniform magnetic field $B_0$ there are only two
decay processes kinematically allowed, $\gamma_{\|}\to\gamma_\perp+
\gamma_\perp$ and $\gamma_{\|}\to\gamma_{\|}+\gamma_{\perp}$ \cite{2}.
Here the subscripts $\perp$ and $\|$ will denote polarizations of the photon
with respect to the vector $B_0$.
More precisely, in  presence of a magnetic field
one has a distinctive plane, namely the $kB_0$ plane. 
One takes the linear polarization of the
magnetic field of the photon parallel  and orthogonal  to
this plane as the two independent polarizations of the photon, $\|$ and
$\perp$, respectively.

The vacuum in the presence of the field $B_0$ acquires an index
of refraction $n$, and the photon dispersion relation is modified from
$k/\omega=1$ to $k/\omega=n$. The indices of refraction $n_{\|,\perp}$
can be calculated from the Heisenberg--Euler effective lagrangian.
Adler showed that for subcritical fields in the limit
of weak vacuum dispersion only the splitting mode $\|\to\perp+\perp$
operates below pair production threshold.
For weak dispersion $n_\perp=1+{7\over90}\,\beta$ and $n_{\|}=1+{2
\over45}\,\beta$, where $\beta={e^4h\over
m^4c^7}\,B^2_0\sin^2\theta$ and $\theta$ is the angle  between 
$k$ and $B_0$. It is mentioned by Harding {\it et al.} that in
magnetar models of soft gamma repeaters \cite{8}, where supercritical
fields are employed,
moderate vacuum dispersion arises. In such a
regime, it is not clear whether Adler's selection rules still endure
since in his analysis higher order contributions to the vacuum
polarization are omitted. In \cite{8} photon cascades are considered for
the case where all three photon splitting modes allowed by CP
invariance are operating. Baier {\it et al.} \cite{5} have found
 that there
is only one allowed transition $(\|\to\perp+\perp)$
 for any magnetic field.
They suggested that a photon cascade could develop only if magnetic
field changes its direction. It seems that the question on the validity
of Adler's rule for a non weak vacuum dispersion deserves a further
study. In this work we consider photon cascades when both kinematically
allowed modes $(\|\to\perp+\perp$ and $\|\to\|+\perp)$ operate.

The interaction operator for the decay $\| \to \perp+\perp$
 is known to be
\cite{3}
\be
                                  \label{2}
V_1(t)=\lambda_1\int(B_0E_1)(B_0E_2)(B_0B)d^3x,
\ee
where the coupling constant $\lambda_1=13e^6/315\pi^2m^8$ and
magnetic and electric parts of photon field are
\be
                     \label{3}
B=i(4\pi)^{1/2}k\times E_{\|}e^{-i(kr-
\omega t)}a_{\|}(k),~~
E_1=-i(4\pi)^{1/2}\omega_1E_{\perp} e^{i(k_1
r-\omega_1t)}a^+_{\perp}(k_1)
\ee
and similarly for $E_2$, here $\omega= \omega_{\|}(k)$ and 
$\omega_i= \omega_{\perp}(k_i)$, $i=1,2$. Here $B$,
$B_0$, $E_i$ are three-dimensional fields and $k_i$
are three dimensional vectors.

For the decay $\|\to\|+\perp$ one has a similar
interaction operator with the operator structure
\be
                       \label{6}
{\cal A}^{+}(t)=\lambda a^{+}
_{\|}(p-k)a^+_{\perp}(k)a_{\|}(p) e^{-itE}
\ee 
where
$E=\omega_{\|}(p)-\omega_{\|}(p-k)-\omega_{\perp}(k).
$
The coupling constant $\lambda$ in this case can be estimated as
$\lambda
/\lambda_1=\alpha(B_0/B_{cr})^2$, where $\alpha$ is
the fine structure constant, $\alpha=e^2/\hbar c$ and $B_{cr}=m^2
c^3/ e\hbar\,\simeq4.4\times10^{13}$ Gauss.

Let us consider a photon cascade created by a photon with momentum
$p$ and polarization $\|$. The photon splits as
$\gamma_{\|}(p)\to\gamma_{\perp}(k_1)+\gamma_{\|}(p-k_1)$.
Then one has the next generation of splitting:
$\gamma_{\|}(p-k_1)\to\gamma_{\perp}(k_2)+
\gamma_{\|}(p-k_1-k_2)$ etc.
After $N$ generations of splitting one gets a cascade with $N$
photons with $\perp $ polarization and momenta
$k_1, k_2\dots $, $k_{N}$ and also one photon with
$\|$ polarization and momentum
${{k}}-
k_1-\dots-k_N$.
An example of a cascade with two generations is shown in Fig.2.
\begin{figure}
\begin{center}
\epsfig{file=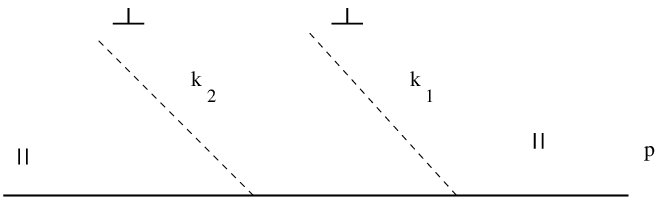,
   width=350pt,
  angle=0
 }

\end{center}
\label{Fig2}
\caption{Cascade: $\gamma_{\|}(p)\to\gamma_{\perp}(k_1)+\gamma_{\|}(p-k_1)
\to \gamma_{\perp}(k_1)+\gamma_{\perp}(k_2)+\gamma_{\|}(p-k_1-k_2)$}
\end{figure}

Our goal is to consider cascades with real
photons (i.e. on the mass shell)
including the intermediate states (compare with what one sees in
the Wilson camera). These cascades can been drawn as
shown in Fig.2. This diagram
is not a Feynman one because all the lines (including an
intermediate one) correspond to
real particles on the mass shell and not to virtual states.
More precisely all the lines in the diagram
are ``dressed'' lines on the mass shell and
the initial photon $\gamma_{\|}(k)$ is prepared in a special way
such that it undergoes the decay in a finite time.  So we cannot
use the standard $S$-matrix approach and the standard Feynman diagram
technique to describe this process.
The diagram is also not a diagram in the
non--covariant diagram technique \cite {CDG} because we have 
the conservation
of energy at every vertex.  The cascade in Fig.2 may be 
intuitively described
by the following state
\be 
\label{8} 
|\psi( p, k_1,k_2)\rangle=f(p,k_1)f(p-k_1,k_2)  
a^+_{\|}(p-k_1-k_2) 
a^+_\perp(k_2)  
a^+_\perp(k_1)
|0\rangle
\ee
where momentum conservation
is built in creation and annihilation operators and energy conservation is
accounted for by the factor $f(p,k)= 
f(\omega_{\|}( p)-\omega_\perp(k)-\omega_{\|}(p- k)) $ where
$f(\omega)$ is a function with  support at $\omega=0$.  As we shall see
below this is not the $\delta$--function but roughly speaking its "square
root".  Indeed, the transition amplitude between two cascade states
is given by scalar product
\be
\label{9}
\langle\psi(p',k'_1,k'_2)|\psi(p, k_1,k_2)\rangle=
|f(p,k_1)|^2|f(p-k_1,k_2)|^2
\delta(p- p')
\delta(k_1-k'_1)\delta(k_2-k'_2)
\ee
\begin{figure}
\begin{center}
\epsfig{file=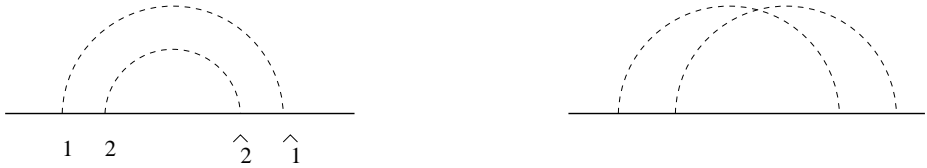,
   width=350pt,
  angle=0
 }

\end{center}
\label{Fig3}
\caption{Non-crossing and crossing diagrams}
\end{figure}

Notice that in the scalar product (\ref{9})
only the non--crossing diagram Fig.3a contributes. In fact the
contribution from the crossing diagram in Fig.3b is negligible 
because of
conservation of energy and momentum.  This is the crucial point where
the difference between our diagrams describing real particles
in intermediate states and the Feynman diagrams having virtual particles
in intermediate states is manifested. In the Feynman diagram
technique the amplitude
of emission of the two photons is represented by a sum of two diagrams
differing by the order in which the two photons are emitted. Here we have
only one diagram, Fig.2.

Now let us observe that if in (\ref{8}) we replace the operators
$a^+_\perp(k_1)$ and $a^+_\perp(k_2)$ by
the quantum
Boltzmann operators $b_\perp(k_1)$ and $b_\perp(k_2)$
satisfying the relations (\ref{i1}),
i.e.  
$b_\perp(k)b^+_\perp( p)=\delta(k-p)$,
then it will be
automatically guaranteed that only the non-crossing diagrams survive.
Therefore it is natural to describe cascade wave functions in terms 
of these operators.  
It is well known that standard free photons are bosons.
Therefore to see the quantum Boltzmann statistics we have to prepare
a special state depending on the interaction.
In fact it is natural to expect
that the cascades with physical intermediate states occur at a time scale
slower than the one occurring in the standard $S$-matrix approach to
multiparticle production. A natural method, leading to this result, is
suggested by the stochastic limit technique.

Now let us consider the question how one can prepare a state
which exhibits the new statistics for photons in cascade. If we would deal
with the scattering of 2--particle states at
infinite time ($S$--matrix) we simply have to consider two Feynman diagrams
to take into account the Bose statistics of photons.  
However in the cascade
we deal with evolution in finite time and the states of photons
$\gamma_\perp(k_1)$ and $\gamma_\perp(k_2)$ are prepared in a
special way because they are emitted at times $t_1$ and $t_2$, respectively.
Therefore, there is a reason not to add the second diagram. 
A special procedure which is adequate to this situation is the
stochastic limit technique described in Sect.3.
In our case the master field is given by the asymptotic behaviour of the
following collective operator

\be 
\label{11c} 
\lim_{\lambda\to 0}
\frac{1}{\lambda} a^{+}_{\|}(p-k)a^+_{\perp}(k)a_{\|}
(p) e^{-itE/\lambda^2}=
{\cal B}^{+}(p,k)  
 \ee 
where $E$ is the same as in (\ref{6})
and as in Sect.3 -- Sect.5 the limit is meant in the sense
of the Wightman correlation functions. 

As it follows from Theorem 1 
 the master field ${\cal B}^{\pm}(p,k)$ satisfies
the following commutation relations
\be
                                     \label{14}
{\cal B}(p,k,t)
{\cal B}^+(p',k',t')=
2\pi\delta(t-t')\delta(E)\delta(k-k')\delta(p-p')
b^{+}_{\|}(k)b_{\|}( k)
\ee
and $b_{\|}(p) , b^+_{\|}(p')$ satisfy the  
 relations
\be
                                   \label{13}
b_{\|}(p) b^+_{\|}(p')=\delta(p-p')
\ee

The presence of the $\delta (E)$-factor 
($E=\omega_{\|}( p)-\omega_{\|}(p-k)-\omega_{\perp} (k)$
) has two important physical
consequences. First,
the commutation relations for the $B^\#$  are not  a consequence of
the corresponding relations for $b^{\#}_{\|}$  and $b^{\#}_{\perp}$:
the three photons are entangled into a single new object
(triphon). Second, the triphon creation and annihilation operators
$B^{\#}$  operate not on the usual Fock space but in
interacting Fock space.

By introducing the auxiliary creation and annihilation operators
$b(t),b^+(t)$, $b_{\perp}(p) , b^+_{\perp}(p)$,
satisfying the quantum Boltzmann relations
\be
b(t)b^+(t')=\delta(t-t'),
\label{bb}
\ee
\be
b_{\perp}(p) b^+_{\perp}(p')=\delta(p-p')
\label{bpb}
\ee
and introducing the symbolic relation
\be
                                   \label{13a}
{\cal B}^{+}(p,k,t) =
b^+(t)b^+_{\|}(k_1) b_{\perp}^+(k_2) b_{\|}(k)(2\pi)^{1/2}
\delta_{1/2}(E)
\ee
we can disentangle the master field by expressing it as a product of
individual Boltzmannian fields. Here the notation $\delta_{1/2}(E) $
is purely symbolic and it simply means that, since
the new commutation relation (\ref{14}) is quadratic in
the master creation and annihilation operators, if in the
above symbolic relation we consider the symbol 
$\delta_{1/2}(E) $ as a scalar
function satisfying the formal relation $[\delta_{1/2}(E)]^2=\delta (E) $,
and if we use the right hand side of (\ref{13a}) to express the left
hand side of (\ref{14}), then the standard Boltzmannian relations 
(\ref{bpb}) and (\ref{13})
 will reduce (\ref{14}) to an identity.
The intuitive understanding of the "disentangling" relation 
(\ref{13a}) is that the
triphon master field ${\cal B}^{+}(p,k,t)$ can be
expressed as the  product of three "Boltzmannian photons": one can think
that each photon of the triple has its own (Boltzmannian) creation and
annihilation operators depending on its own momentum, however
in the master field these operators can only appear in the combination
given by the right hand side of (\ref{13a}) 
and there is a constraint among the three
momenta expressed by energy conservation.

Notice the Boltzmannian white noise
relation (\ref{bb}), which makes our model particularly suitable for Monte
Carlo simulations.
The origin for these new commutation relations lies in
the fact that the crossing diagrams in the computation of 
the matrix element
(\ref{8}) are suppressed in the weak coupling/large time limit.

A photon splits into two not only in a magnetic field
but also in a nonlinear medium. In fact such processes are well known
in nonlinear quantum optics, see for example \cite{WM}. In the
nonlinear process of parametric down conversion a high frequency photon
splits into two photons with frequencies such that their sum 
equals that of
the high-energy photon.  The two photons  produced in this
process possess quantum correlations and have identical intensity
fluctuations.

In conclusion, in this paper we have argued that photon cascades in a
strong magnetic field might create a new type of entangled states
(triphons) which obey not Bose but the quantum Boltzmann statistics. 
This
prediction is based on the assumption that both kinematically allowed
photon splitting modes operate.

Given the validity of this assumption we prove that, in the stochastic
regime the intermediate photons in a cascade are real and virtual
particles. The dominating contributions to the dynamics come from these
entangled triples of photons  which behave like single new particles
(triphons) whose statistics can be experimentally observed by counting
the emitted photons in the corresponding cascades.

As explained in the introduction, the time scale in which our
predictions are true is long compared to the strength of the coupling
but short if compared to the time scale of the $S$--matrix approach.
This remark should be kept into account in a possible experimental
verification of these predictions.
A better
theoretical understanding of the photon splitting with a non weak 
dispersion
is required.  From the experimental side new more precise devices such
as the planned Integral mission \cite{7} might significantly advance our
understanding of the fundamental problem of photon statistics.

{\bf Acknowledgments.} I.Ya.A. and I.V.V. are grateful 
to the Centro Vito Volterra
Universita di Roma Tor Vergata for the kind hospitality. This work is
supported in part by INTAS grant 96-0698,  I.Ya.A. 
 is supported also by RFFI-99-01-00166
and I.V.V. by RFFI-99-01-00105

\end{document}